# Hybrid Online Certificate Status Protocol with Certificate Revocation List for Smart Grid Public Key Infrastructure

**Hong-Sheng Huang[1,*], Zhe-Yi Jiang[1,†], Hsuan-Tung Chen[2], Hung-Min Sun[3]**
[1*,1†] Institute of Information Security, National Tsing Hua University, Hsinchu, Taiwan
[1]E-mail address: ray.h.s.huang@m111.nthu.edu.tw
[1,†]E-mail address: zy.jiang@m111.nthu.edu.tw

[2] Information and Communications Research Laboratories, Industrial Technology Research Institute, Hsinchu, Taiwan
E-mail address: hsuantung@itri.org.tw

[3] Department of Computer Science, National Tsing Hua University, Hsinchu, Taiwan
E-mail address: hmsun@cs.nthu.edu.tw

**Abstract**

Hsu et al. (2022) proposed a cryptographic scheme within the public key infrastructure to bolster the security of smart grid meters. Their proposal involved developing the Certificate Management over CMS mechanism to establish Simple Certificate Enrollment Protocol and Enrollment over Secure Transport protocol. Additionally, they implemented Online Certificate Status Protocol (OCSP) services to independently query the status of certificates. However, their implementation featured a single OCSP server handling all query requests. Considering the typical scenario in smart grid PKI environments with over tens of thousands of end-meters, we introduced a Hybrid Online Certificate Status Protocol mechanism. This approach decreases demand of query resources from the client to OCSP servers collaborating with Certificate Revocation Lists. Our simulations, mimicking meter behavior, demonstrated increased efficiency, creating a more robust architecture tailored to the smart grid meter landscape.

## 1. Background/ Objectives and Goals

The integration of public key infrastructure within smart grids, as expounded upon by Hsu et al. [1], presents a significant augmentation in security measures. Leveraging protocols such as RFC 8894 SCEP[2] and RFC 7030 EST[3], this infrastructure ensures the secure transmission of data between smart meters and servers, guaranteeing confidentiality, integrity, and non-repudiation. Notably, the functionalities of RFC 6960 OCSP [4] services have been extended to an autonomous server, distinct from the Certificate Authority (CA) depicted in Figure 1. However, the current configuration assigns the OCSP server the task of handling all query requests, inclusive of those from the smart grid and diverse meter manufacturers.

In the context of a smart grid environment housing over tens of thousands of smart meters, as acknowledged within Hsu's study, each meter manufacturer is mandated to oversee all meters, as depicted in Figure 2. Consequently, an independent OCSP server assumes the responsibility of querying the certificates of all meters whenever a client initiates OCSP requests. This operational setup necessitates querying the entire database to ascertain the status of a single certificate, leading to considerable resource depletion.

However, within the framework of the certificate management system, a pivotal component, the Certificate Revocation List (CRL), serves as a compilation of digital certificates encompassing all revoked status certificates from the certificate database. Traditionally, the CRL serves the purpose of verifying the identities of users, computers, or entities within a networked environment. In our scenario, it validates the identities of smart meters. Typically overseen by the CA, the CRL is generated and signed using the root CA certificate, thus ensuring its reliability for other entities. Its structural representation aligns closely with Figure 3.

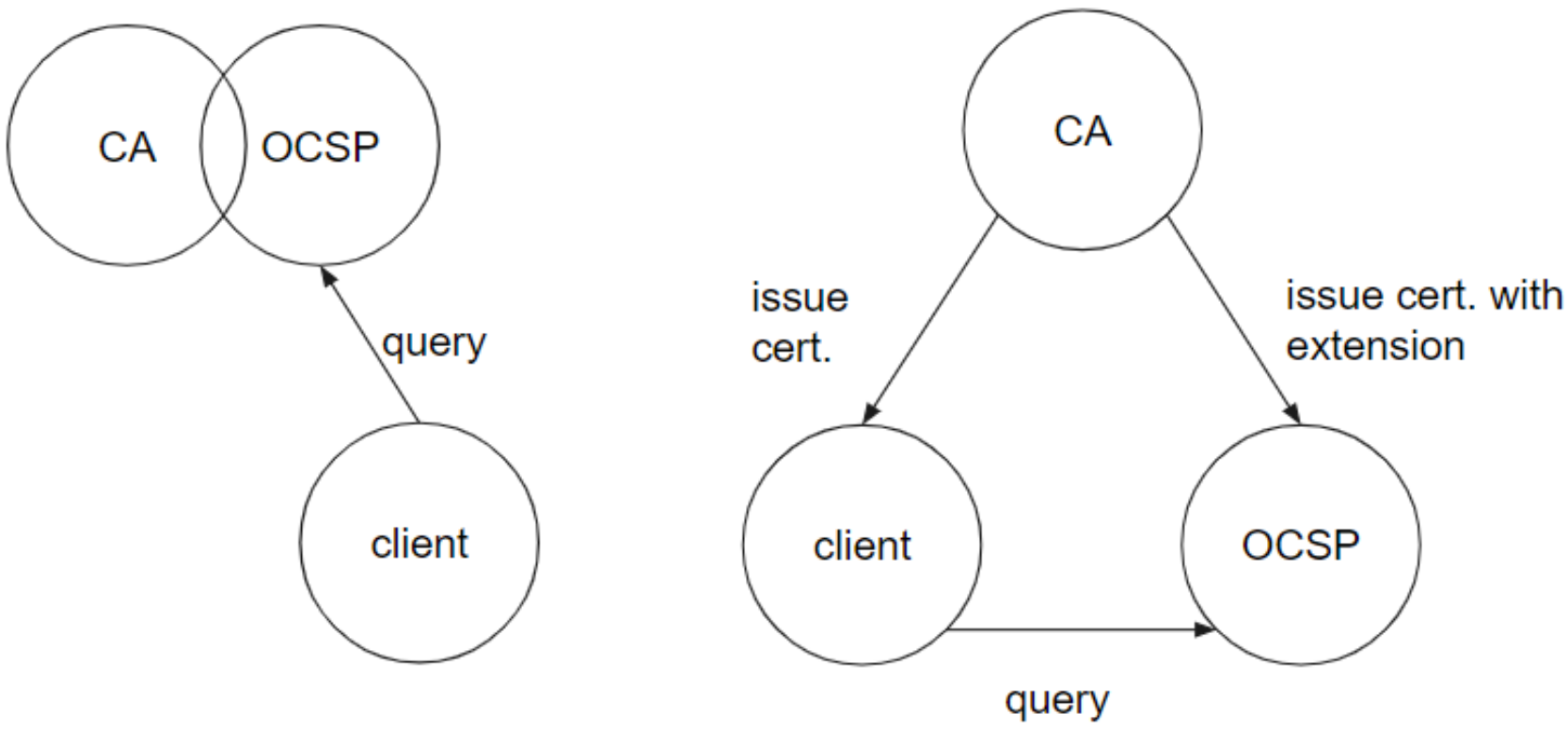

Figure 1. OCSP type

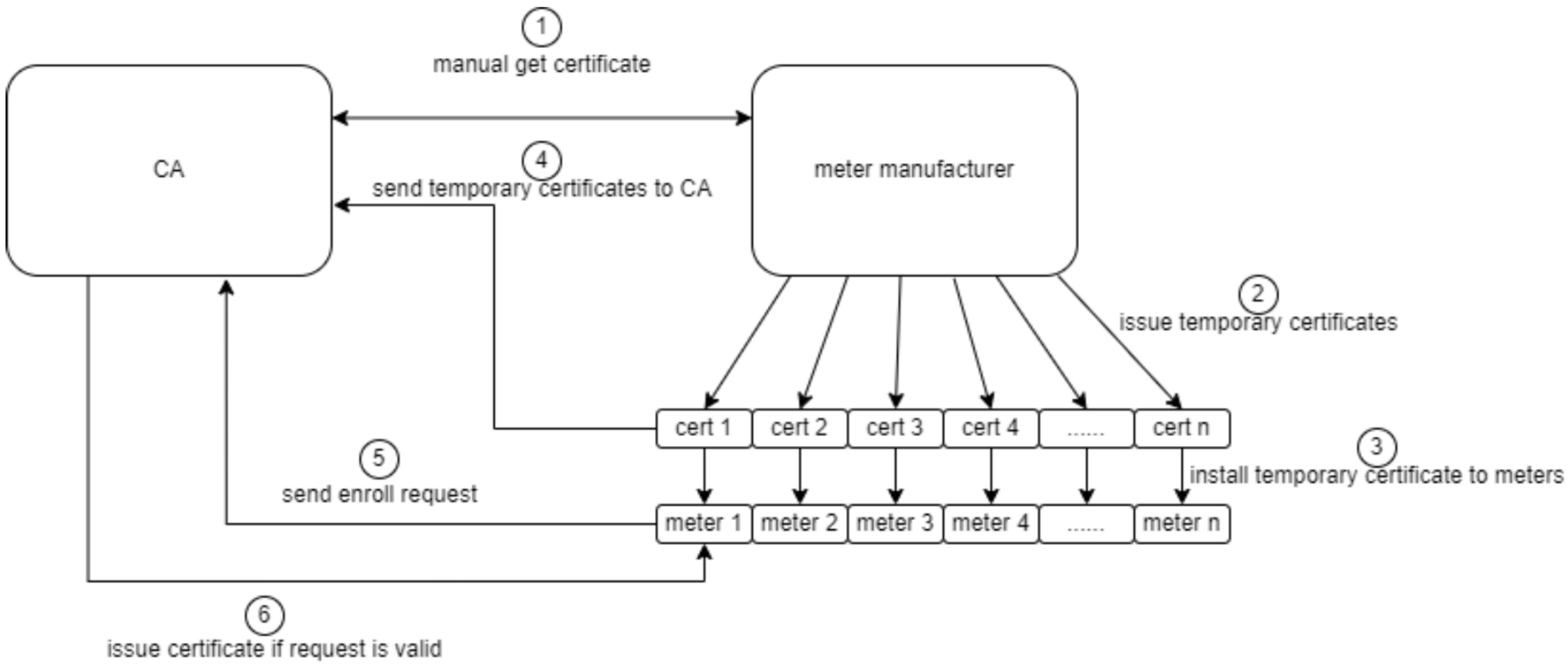

Figure 2. Certificate whitelist authentication method [1]

To optimize computational resource utilization, our initiative involves the development and upkeep of a CRL dedicated to storing revocation certificates, thereby establishing an efficient blacklist. Integrating this CRL within the OCSP server framework facilitates the creation of a robust blacklist database, ensuring High Availability (HA).

When an OCSP request is received from a client, the system performs a query against the blacklist database. Certificates listed within the CRL are flagged as invalid, indicating their inclusion in the blacklist. Conversely, certificates absent from the CRL are considered valid. Additionally, in scenarios where the OCSP server is unable to respond to a query request, the CRL serves as a backup, enabling clients to independently verify the certificate status.

This strategic integration empowers the OCSP to diligently identify revoked certificates, thereby effectively conserving computational resources.

```
user@user-Standard-PC-i440FX-PIIX-1996:~/smart_grid_stuff_alt/crl$ cat crl.str
Certificate Revocation List (CRL):
        Version 1 (0x0)
        Signature Algorithm: sha256WithRSAEncryption
        Issuer: C=aa, ST=aa, L=aa, O=aa, OU=aa, CN=rootca
        Last Update: May  4 19:57:27 2023 GMT
        Next Update: NONE
Revoked Certificates:
    Serial Number: 221A0A99711F9968
        Revocation Date: May  4 19:57:27 2023 GMT
        CRL entry extensions:
            X509v3 CRL Reason Code:
                Key Compromise
    Serial Number: 308C707EA89F47A5
        Revocation Date: May  4 19:57:27 2023 GMT
        CRL entry extensions:
            X509v3 CRL Reason Code:
                Key Compromise
    Serial Number: 5238F3475665F7C4
        Revocation Date: May  4 19:57:27 2023 GMT
        CRL entry extensions:
            X509v3 CRL Reason Code:
                Key Compromise
    Signature Algorithm: sha256WithRSAEncryption
    Signature Value:
        9b:df:66:e2:d6:7b:82:41:ab:39:53:e0:a5:69:5b:07:6a:90:
        98:7e:bc:30:bf:4e:5f:2d:38:c8:fa:2c:09:43:62:4f:3d:38:
        65:8c:2b:2c:b2:a5:6e:3f:a1:18:a9:c1:04:64:65:13:77:3b:
        e4:8f:93:f9:fb:13:82:34:20:c6:b5:e4:94:1f:51:b5:f5:11:
        76:88:a9:99:d4:92:ca:3c:7a:6a:56:2b:8c:c7:a6:57:ce:42:
        50:47:2e:a4:58:63:3a:58:86:9b:69:9f:66:a2:82:f8:9b:a1:
        f5:fa:72:72:6e:d7:48:62:a5:45:db:c9:88:88:49:89:31:b6:
        ed:77:9c:8a:f2:e6:ac:26:be:cb:eb:e5:c4:7b:da:9c:ae:b1:
        9e:8d:f6:0c:cc:a3:6c:87:5f:0b:f9:cc:f6:9f:6c:c9:68:82:
        d4:9c:63:7a:67:c5:9f:10:a6:88:64:5c:ad:1b:50:b1:79:06:
        dc:d9:d2:74:b0:c3:8a:a7:c5:2e:e9:75:aa:08:02:51:fa:c9:
        1c:2f:9e:41:37:fb:58:15:1e:eb:d2:bd:c8:55:c0:c1:9b:0f:
        ba:81:fe:08:12:4c:c5:d9:d1:10:53:75:a0:8b:6e:db:c6:63:
        0e:d9:c0:59:c0:a7:b4:7f:12:bc:0b:ff:12:9d:ef:ea:af:95:
        7d:5d:27:fa
```

Figure 3. Certificate Revocation List

## 2. Methods

### 2.1 Hybrid Online Certificate Status Protocol

To streamline resource utilization, we've integrated the Certificate Revocation List (CRL) with the Online Certificate Status Protocol (OCSP), as depicted in Figure 4. This amalgamation transforms the CRL into a blacklist housing revoked certificates. We've established a schedule for the CRL to update every hour, ensuring the blacklist remains current. Subsequently, the OCSP examines the CRL contents and stores all revoked certificates in its database.

Following above setup, when the OCSP server receives query requests, two scenarios unfold:

1. If the requested certificate is found in the CRL, it indicates the certificate is invalid.
2. If the requested certificate isn't present in the CRL, it signifies the certificate remains valid.

Upon verifying the certificate's status, the OCSP server promptly responds to the client. As a result, the OCSP server no longer needs to search the entire certificate database for each query request, thus conserving computational resources that can be allocated to other services.

We advocate the adoption of "Hybrid OCSP," defined as a mechanism that strategically integrates the advantages of both traditional OCSP and CRL. This approach seeks to harness the strengths of both mechanisms to enhance the efficiency and reliability of certificate validation processes. In instances where the OCSP server experiences downtime or becomes overloaded, necessitating a backup service, our proposed solution involves the utilization of CRL as a reliable contingency measure.

The OCSP is specifically designed for the verification of signed certificates. In situations where an endpoint device requires the verification of a large volume of certificates, it is advisable to employ CRL instead.

This strategic integration of Hybrid OCSP aims to ensure a resilient and uninterrupted certificate validation infrastructure, thereby addressing potential challenges associated with server reliability and load management.

### 2.2 Scenario Simulation

In the client simulation, we emulate a comprehensive smart grid environment, as depicted in Figure 4, consisting of endpoint devices functioning as clients, a Certificate Authority (CA) server, a Registration Authority (RA) server, and an Online Certificate Status Protocol (OCSP) server, and notably, the Certificate Revocation List (CRL) hosted on the CA server.

Clients employ factory-issued certificates and communicate with the RA server via a certificate management protocol like Figure 2. Subsequently, the RA server interacts with the CA server to perform necessary operations and returns the finalized certificate to the client.

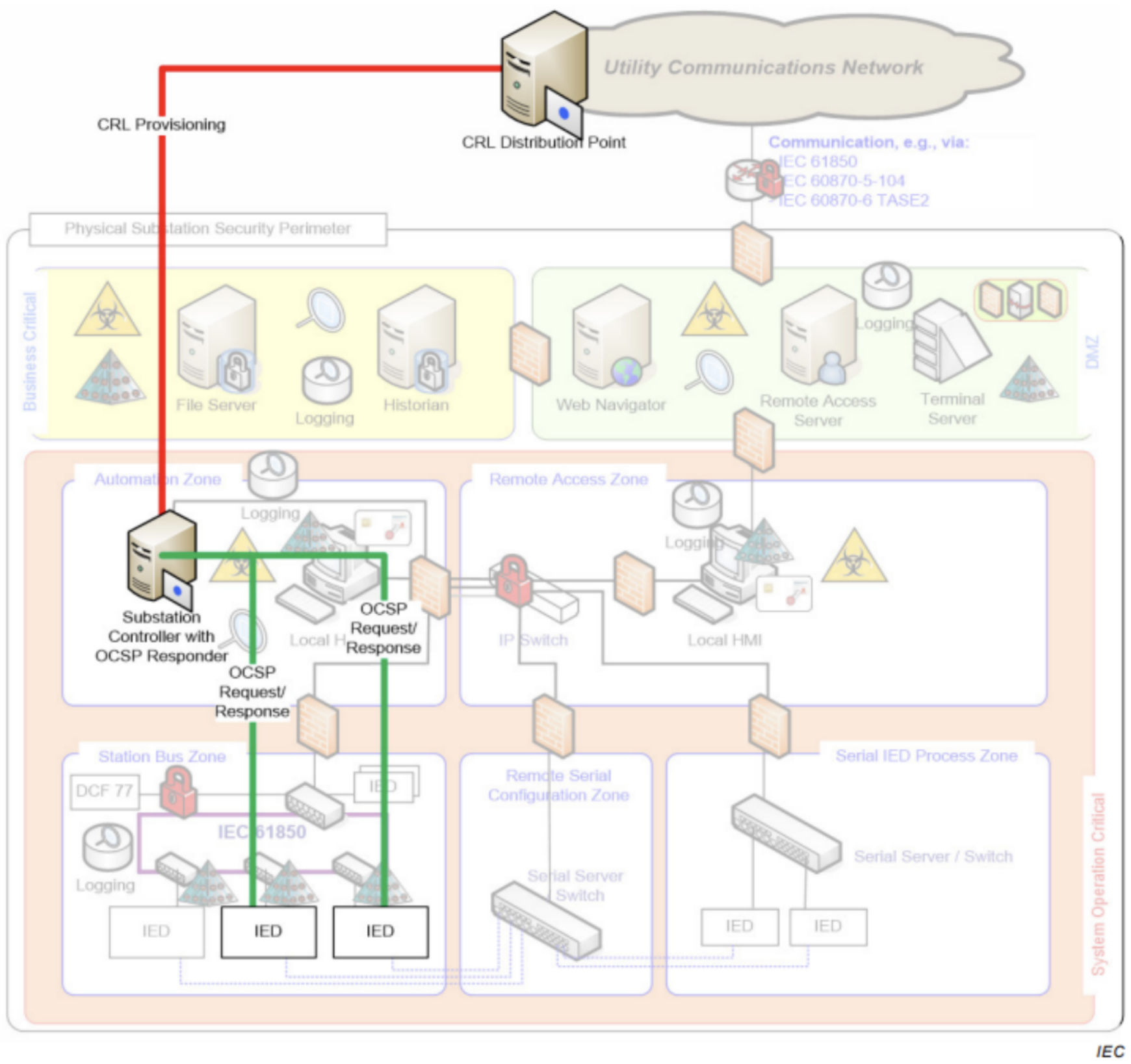

Figure 4. Hybrid OCSP architecture[6]

The operational workflow involves the CA server routinely generating the CRL list, which is also routinely downloaded by the OCSP server. The OCSP server parses the CRL list, extracting and storing serial numbers in its database. Upon receiving an OCSP request, the server conducts a database search to determine the existence of the serial number. If the serial number is found, the server responds with a status of "revoked"; otherwise, it responds with a status of "good" to the client.

In circumstances where the OCSP service experiences an outage, endpoint devices exhibit the capacity to seamlessly pivot towards utilizing CRL. As previously highlighted, clients retain the autonomy to independently download CRL from distributed certificate repositories. This approach ensures clients maintain continuous access to certificate status verification within the dynamic framework of a smart grid environment.

## 3. Results

In this section, we investigated the data consumption associated with Certificate Revocation Lists (CRL) and Online Certificate Status Protocol (OCSP) on our experimental device. The device specifications include host operating system, Proxmox VE, equipped with Intel 10980EX CPU, and guest operating system, Ubuntu 20.04.3 LTS, with 11 GB of RAM.

CRLs exist in two primary formats: DER format, which is stored as raw binary, and PEM format, which entails raw binary with base64 encoding. Generally, CRLs in PEM format exhibit larger sizes than those in DER format.

We aimed to quantify the network data associated with two methods: OCSP requests via OpenSSL and CRL downloads via Wget. Our focus encompassed three scenarios: OCSP, CRL PEM format, and CRL DER format, all pertaining to RSA certificates.

### 3.1 An Empirical Analysis of Credential Revocation Mechanisms: Assessing Network Data Consumption and Endpoint Device Limitations

The findings presented in Table 1 and illustrated in Figures 5, 6, and 7 indicate, if the CRL in PEM format consistently exceeds 14 records or the CRL in DER format consistently surpasses 24 records, we recommend prioritizing the utilization of OCSP. This recommendation is founded on two primary considerations. Firstly, as illustrated above, OCSP demonstrates advantages in terms of network data efficiency. Secondly, consideration must be given to the inherent limitations of endpoint devices, characterized by relatively weak CPU performance and limited storage capacity[5]. Specifically, in instances where a CRL is extensive, there exists a risk that endpoint devices may struggle to download and decode the entire CRL within a constrained timeframe.

Table 1: Comparison of the scenarios

| | Aggregate Data (req+resp) | Num of CRL records |
|---|---|---|
| OCSP | 1,841 bytes | x |
| CRL PEM format | 1,830 bytes | 14 |
| CRL DER format | 1,851 bytes | 24 |

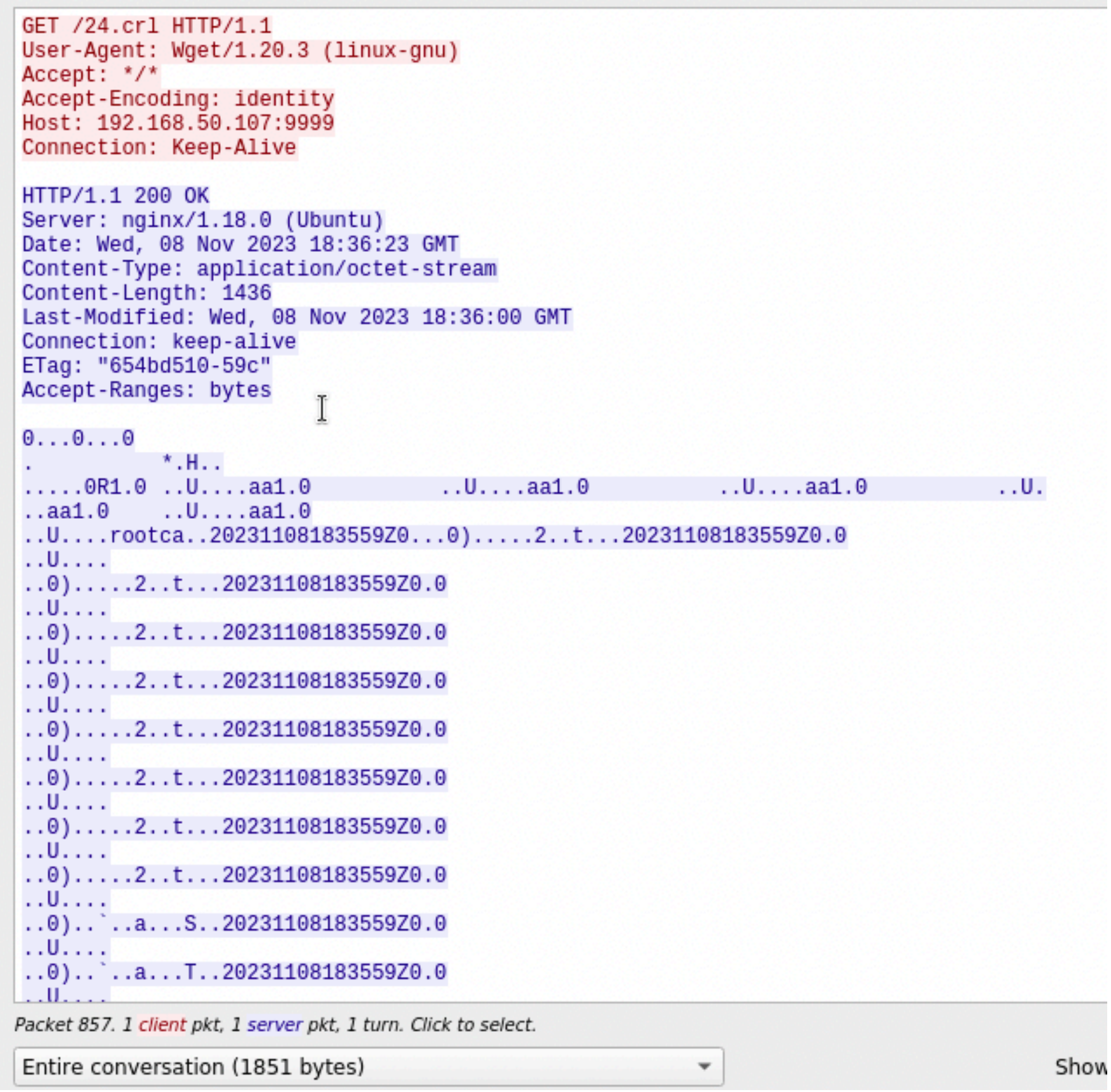

Figure 5. Wireshark CRL(raw binary) data

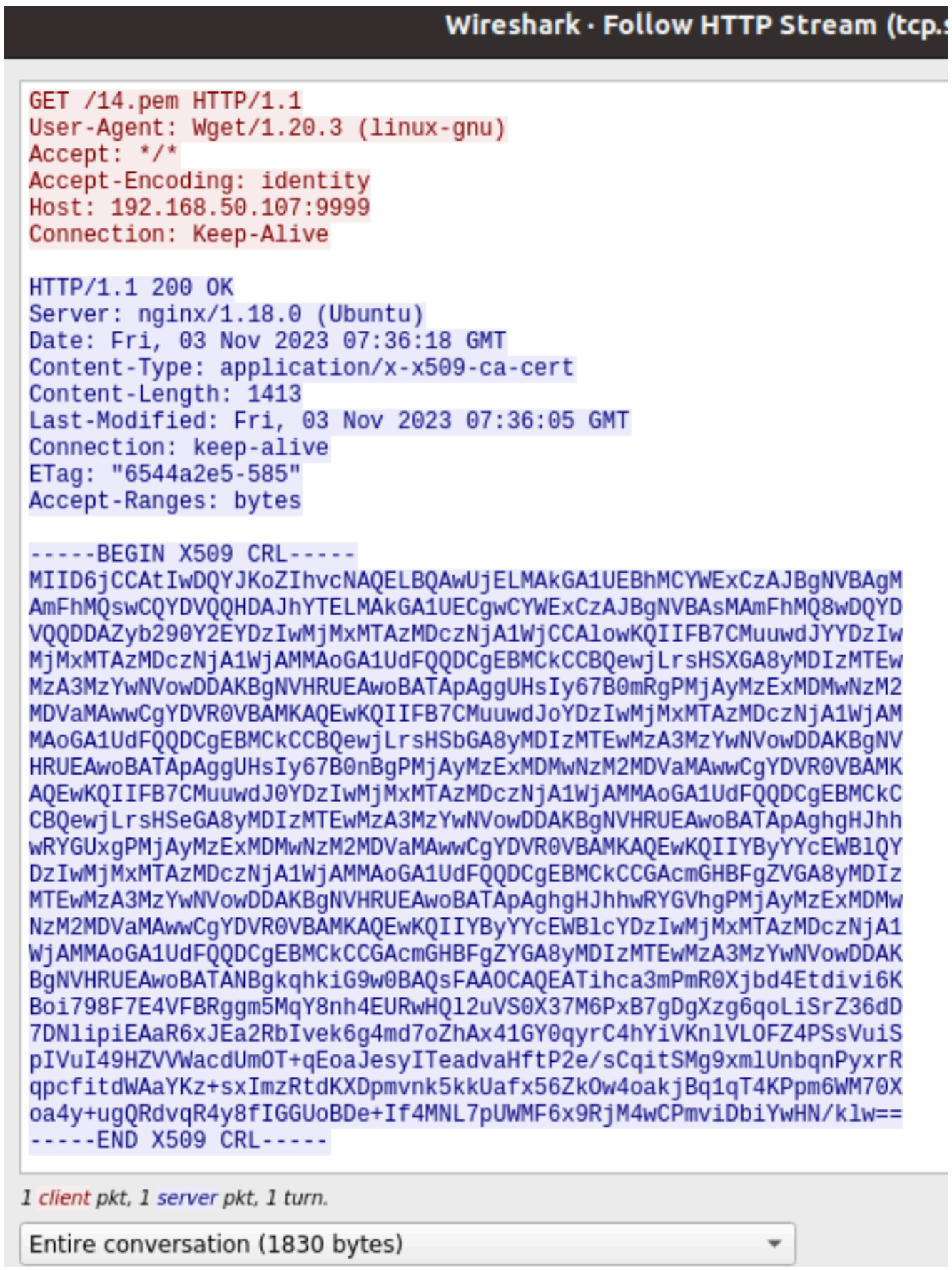

Figure 6. Wireshark CRL(base64) data

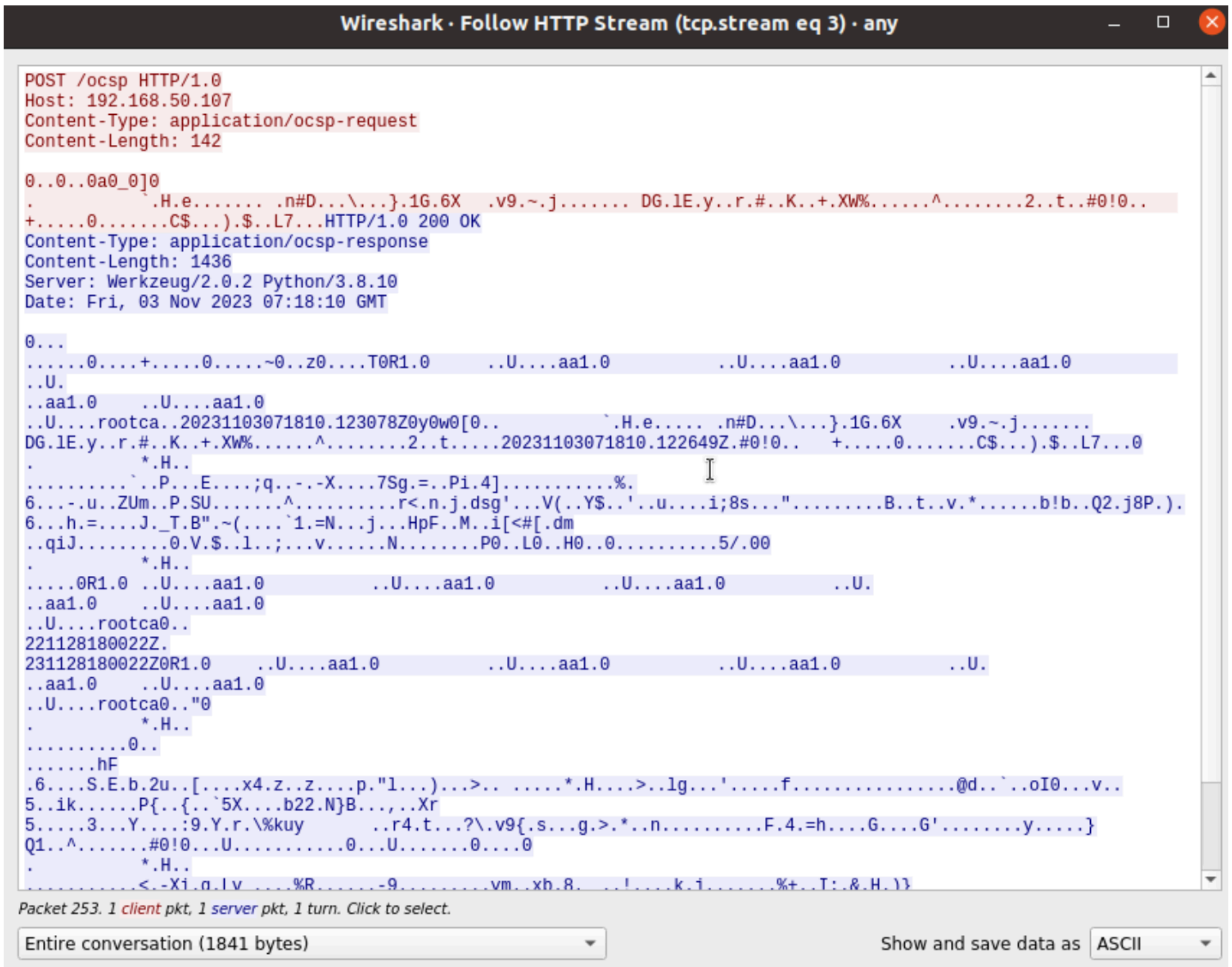

Figure 7. Wireshark OCSP data

### 3.2 Hybrid OCSP Server Benchmark

We conducted a comprehensive benchmarking analysis of our implemented OCSP server, which was developed using the Python programming language with the Flask framework.

We performed a summation of the elapsed time values and subsequently computed the average of these time intervals, as delineated in Table 2.

The performance of the Hybrid OCSP server was deemed acceptable within the current configuration. Considering potential avenues for performance improvement, we posit that augmenting the computational resources by increasing the number of CPU cores and transitioning to an Ahead-of-Time compilation approach could yield enhanced results.

Table 2: Hybrid OCSP server benchmark

| | Time consumed | Avg. request time |
|---|---|---|
| 100 thousand OCSP requests on 4 cores cpu | approxi. 17 mins | 0.0102s /per request |
| 1 thousand OCSP requests on 2 cores cpu | approxi. 30 secs | 0.029s /per request |

### 3.3 Conclusion

The primary objective of introducing Hybrid Online Certificate Status Protocol (OCSP) is to address the prevalent issue of OCSP servers encountering a high volume of OCSP requests. In response to this challenge, we propose a novel approach termed "Hybrid OCSP," which strategically leverages the strengths of both traditional OCSP and Certificate Revocation Lists (CRL). In contemplating future developments, there remain areas for enhancement. Notably, the current Hybrid OCSP server, implemented in python, exhibits suboptimal performance in comparison to C.

In summary, we have successfully implemented Hybrid OCSP within the context of a smart grid Public Key Infrastructure (PKI) mechanism for endpoint devices. This innovative solution enables the endpoint device to dynamically select the protocol—either traditional OCSP or CRL—for verifying the validity of its certificate. The benchmarking results underscore the acceptability of the performance achieved by the Hybrid OCSP implementation. Furthermore, we elucidate the rationale behind our recommendation for smart grid endpoint devices to preferentially adopt Hybrid OCSP , thereby contributing to a more resilient and efficient PKI infrastructure.

### 3.4 Acknowledgments

We extend our heartfelt gratitude to Mr. Hsuan-Tung Cheng, whose invaluable assistance proved instrumental in resolving numerous complex issues. His guidance and unwavering support were pivotal in navigating challenging situations, and his dedication in tackling difficult problems alongside us was truly commendable.

Additionally, we wish to express our sincere appreciation to Professor Hung-Min Sun for his consistent support and mentorship, which has been immensely valuable to us throughout our endeavors.

This work was supported in part by the Ministry of Science and Technology, Taiwan, under Project MOST-110-2221-E-007-040-MY3 and MOST-111-2221- E-007-078-MY3.